\documentclass[11pt,english]{article}
\usepackage{graphicx}
\usepackage{amsmath}
\usepackage{amssymb}
\usepackage{babel}
\usepackage{ragged2e,amsfonts,graphicx,caption,subcaption,xcolor}
\usepackage[left=2cm,right=2cm,top=3cm,bottom=3cm]{geometry}
\usepackage{hyperref}
\hypersetup{colorlinks=false, linkbordercolor={white}}

\def\laq{~\raise 0.4ex\hbox{$<$}\kern -0.8em\lower 0.62ex\hbox{$\sim$}~}
\def\gaq{~\raise 0.4ex\hbox{$>$}\kern -0.7em\lower 0.62ex\hbox{$\sim$}~}

\def\beq{\begin{equation}}
\def\eeq{\end{equation}}
\def\bea{\begin{eqnarray}}
\def\eea{\end{eqnarray}}

\def \ti {\widetilde}

\def \ra {\rightarrow}
\def \noi {\noindent}

\def\Tr{\mathop{\rm Tr}}

\def \fpu {\dot{\phi}}
\def \fpp {\ddot{\phi}}

\def \fbp {\dot{\fb}}

\def \rb {\overline \rho}
\def \pb {\overline p}
\def \sgb {\overline \sg}

\def \rg {\sqrt{-g}}

\def \fb {\overline \phi}

\def \pa {\partial}
\def \ra {\rightarrow}
\def \ti {\widetilde}
\def \la {\lambda}
\def \ls {\lambda_{\rm s}}
\def \lp {\lambda_{\rm P}}
\def \Ms {M_{\rm s}}
\def \Mp {M_{\rm P}}

\def \da {\delta}

\def \a {\alpha}
\def \ap {\alpha^{\prime}}

\def \Ga {\Gamma}
\def \ga {\gamma}
\def \sg {\sigma}

\def \da {\delta}

\def \r {\rho}

\def \Om {\Omega}
\def \noi {\noindent}

\def \L {{\cal L}_m}

\def \d {{\rm d}}
\def \e {{\rm e}}

\begin{document}
\begin{titlepage}

\thispagestyle{empty}

\begin{flushright}
Preprint BA-TH/805-21\\
\end{flushright}

\vspace{1cm}

\begin{center}
\noindent{\Large \textbf {FROM PRE- TO POST-BIG BANG:\\
\vspace{0.4cm}
AN (ALMOST) SELF-DUAL COSMOLOGICAL HISTORY}}\\

\vspace{1cm}

\noi
\large{M. Gasperini}

\normalsize

\vspace{0,5cm}

{\it Dipartimento di Fisica, Universit\`a di Bari, 
Via G. Amendola 173, 70126 Bari, Italy,\\
and Istituto Nazionale di Fisica Nucleare, Bari, Italy} \\
{e-mail: gasperini@ba.infn.it
 }

\end{center}

\vspace{1cm}

\begin{abstract}
\noindent
 We present a short introduction to a non-standard cosmological scenario motivated by the duality symmetries of string theory,  in which the big bang  singularity is replaced with a "big bounce" at high but finite curvature. The bouncing epoch is prepared by a long (possibly infinitely extended) phase of cosmic evolution, starting from an initial state asymptotically approaching the string perturbative vacuum.
\end{abstract}

\vspace{2 cm}

\centerline{ 
--------------------------------------------}

\vspace{0.5 cm}


\noindent
{\it This paper is dedicated to the memory of {\bf John D. Barrow}, that I met at the ``Fourth Paris Cosmology Colloquium"  (Observatoire de Paris, June 1997). We went to dinner together, and we spent the whole evening talking about cosmology, gravity, and string theory. I was greatly impressed by his uncommon wisdom and modesty, typical of great men. His passing was a tremendous loss to the entire  community of physicists and cosmologists.}


\vspace{0.5 cm}

\centerline{ 
--------------------------------------------}

\end{titlepage}

\tableofcontents

\section{Introduction}
\label{sec1}
\renewcommand{\theequation}{1.\arabic{equation}}
\setcounter{equation}{0}

It is well known that the phase of decelerated expansion predicted by the standard cosmological scenario (see e.g. \cite{1}) cannot be extrapolated back in time to very early epochs without producing a series of (conceptual) problems, like the horizon, flatness, entropy problem, closely related to the properties of the presently observed cosmological state (see e.g. \cite{2}). At early enough times, a phase of accelerated, inflationary expansion is expected to occurr \cite{3,4,5}, in order to avoid such problems.

However, even the inflationary phase -- if characterized by a constant or decreasing behavior of the cosmic curvature scale -- cannot be extended back in time to infinity (or, using the standard terminology, cannot be {\em ``past eternal"}) \cite{6,7,8}: it must have a beginning at a (very early but) finite instant of time.

Before that time our Universe, according to the standard cosmological scenario, is expected to be in an extremely hot, dense and curve primordial state approaching (in a finite time interval) the so called {\em big bang} epoch. Namely, the infinitely curved epoch of the huge cosmic explosion which was the origin of the space-time itself. We can thus say that to the question:  ``How did our Universe begin?", the standard, even inflationary, cosmological model provides the answer: ``Our Universe was born from the initial big bang singularity".

Such a conclusion, however, is based on two important assumptions of the standard scenario: an inflationary phase at decreasing curvature, and the Einstein theory of general relativity, which is a relativistic theory of gravity, but not a quantum theory. Hence, like all classical theories, it has a limited range of validity: the action $S=Et$ of any described physical process has to be much larger than the elementary (Planck) {\em quantum of action}, i.e. $S \gg h$.

In a cosmological context, if we take a typical spatial volume of  size controlled by the Hubble radius $c/H$, we can estimate the total associated energy $E$ by considering the energy density $\r$ of the gravitational sources, determined  by the Einstein equations as $\r \sim c^2 H^2/G$ (where $c$ is the light velocity and $G$ the Newton constant). The corresponding time scale, on the other hand, is provided by the Hubble time $H^{-1}$. By imposing that $S=E t  \sim \r (c/H)^3 H^{-1} \gg h$ we then find the condition:
\beq
 {c^5\over G H^2} \gg h.
 \label{11}
 \eeq
 (This condition, in units $\hbar=c=1$, can also be rephrased as $H \ll \Mp$, where $\Mp$ is the Planck mass).
 
 In the above equation $c$, $G$, $h$ are constant, but the Hubble parameter is time dependent: according to the standard model it grows as we go back in time, and diverges at the time of the big bang singularity. Correspondingly, the ratio $c^5/G H^2$ tends to zero. Hence, before reaching the  big bang epoch, we necessarily enter the regime where the condition (\ref{11}) is violated, and the classical gravitational equations of the standard model are no longer valid. For a reliable description of the primordial Universe we should thus use a more general theoretical approach, compatible with gravity also in the quantum regime. 
 
 A possible candidate for such a theory, able to describe in a consistent way all fundamental interactions, including gravity, at all energy scales and in the quantum regime, is string theory (see e.g. \cite{9,10,11}). Why should we consider string  theory, instead of other alternative approaches to a quantum description of gravity? We have of course many possible answers to this question but, in my opinion, there is a very simple -- and compelling  -- reason: for a quantized string, gravity is not an ``optional" ingredient to be possibly included into the theory; it is an interaction {\em necessarily} predicted by the string model at the quantum level.
 
 Let us briefly recall, in fact, that and extended object like a string corresponds, dynamically, to a constrained system, whose motion has to be consistently described not only in terms of the appropriate Eulero-Lagrange equations, but also imposing a set of constraints to be satisfied at any point along the string trajectory. When the string is quantized such constraints are represented by an infinite (discrete) series of operators (the Virasoro operators), and the Hilbert space of the physical states is formed by (all and only) those states annihilated by the application of the Virasoro operators. And the constraint corresponding to the lowest order operator imposes on the states the mass-shell condition, determining the allowed energy levels of the string spectrum. 
 
 It follows, in particular, that the string states can be ordered as a ``tower" of  growing (and discrete) mass and angular momentum eigenvalues. And if we look at the subset of massless eigenstates we find that the spectrum of the closed strings contains (besides other fields) a symmetric tensor field of rank two, which is transverse and traceless, and which has all the required physical properties to represent a graviton. Thus, quantum string theory must necessarily include a tensor interaction of gravitational type.
 
 But there is more. Indeed, string theory is characterized by the presence of the so-called conformal symmetry, typically associated to the geometry of the two dimensional ``world-sheet" surface spanned by the string during its motion in the external space-time manifold. At the quantum level, the conformal invariance of the free string dynamics plays a crucial role in determining the correct spectrum of the physical string states. Hence, quantum conformal invariance has to be preserved even when a string interacts with any given external field. 
 
 And by imposing the quantum validity of the conformal symmetry (namely, by imposing the absence of ``conformal anomalies") when the string interacts with the tensor field which is a possible candidate to describe gravity, one obtains for such a tensor field a set of differential conditions which are exactly equivalent, at the lowest order, to the Einstein field equations! Hence, the standard gravitational interactions, and even their conventional general relativistic description, must be {\em necessarily} included in the quantum string model of fundamental interactions.
 
 According to string theory, however, the Einstein equations can describe gravity only to lowest order, in the low-energy and small coupling limit \cite{9,10}. At arbitrarily large energy scales the theory also provide (at least in principle) exact gravitational equations, fully consistent with the quantum regime, and they are different from the Einstein equations. The differences are due to the possible presence of additional spatial dimensions, additional components of the gravitational force (e.g. scalar and pseudo-scalar fields), additional higher-derivative contributions to the kinetic part of the action, etc. 
 
 In a cosmological context, in particular, the differences from the Einstein equations turn out to be important when describing the quantum regime of the very early Universe, near the big bang epoch. In such a context, therefore, 
it makes sense to ask the question:``What's new from string theory about cosmology?" and , in particular: ``What's new about the very early epochs at the beginning of our Universe?"


\section{String duality and cosmology}
\label{sec2}

\renewcommand{\theequation}{2.\arabic{equation}}
\setcounter{equation}{0}

In this paper we shall concentrate on the possible cosmological applications of two peculiar aspects of string theory: the duality symmetry and the existence of a fundamental length $\ls$, typical of quantized strings, and related to the string tension (or energy per unit length) $T$ by $\ls^2 = c \hbar/T \equiv 2 \pi \ap$. Let us start with duality. 

Besides the conformal symmetry mentioned in the previous section, another example of genuine ``stringy" symmetry is the so-called target-space duality (or $T$-duality) \cite{12}, typical of quantum closed strings in space-time manifolds with compact spatial dimensions of finite radius $R$. According to this symmetry, the energy spectrum of a string turns out to be invariant with respect to the transformations $R \leftrightarrow  \ls^2/R$ and the simultaneous exchange of its ``winding" modes (depending on the number of times the string is wrapped around the compact dimensions) with its momentum modes (which along  compact directions are characterized by a discrete spectrum, according to standard quantum mechanics).

A cosmological variant of such a symmetry, called ``scale-factor duality" \cite{13,14}, implies that given a cosmological phase characterized by a monotonically decreasing absolute value of the Hubble parameter (hence, decreasing curvature), we can obtain a {\em dual} partner phase characterized by a growing absolute value of the Hubble parameter (and of the curvature), provided we simultaneously perform also a reflection of the cosmic time coordinate $ t \ra -t$. Unlike $T$-duality, there is no need of compact spatial dimensions, and the symmetry is effective also at the level of the tree-level (classical) equations of motion (and not necessarily at the level of the quantum string spectrum).

According to this symmetry, in particular, the present cosmological phase subsequent to the big bang epoch, and well described in the standard model by a space-time geometry with {\em decreasing} $H(t)$, $t>0$, should be preceded   in time by a dual, almost specularly symmetric phase with {\em increasing} $H(-t)$, $t<0$, and occurring {\em before} the big bang (assumed to be localized at $t=0$).

In the above picture of cosmic evolution, both phases are characterized by a Hubble parameter (and a curvature) which diverges as $t \ra 0$. If that would be the case, then the two branches of of the cosmological evolution would be causally disconnected by a singularity, with no chances of merging into a single, coherent models of space-time evolution -- at least, without the explicit introduction of quantum cosmology effects described by the Wheeler-De Witt equation (see e.g. \cite{15} for a recent discussion). It is here, however, that comes into play the other crucial aspect of string theory.

Quantum strings, unlike their classical analogs, have a minimal (or, better, optional) size $\ls$. The physical role of $\ls$ is very similar to the role played for the atom by the Bohr radius, which represents the minimal allowed size of the quantum electronic orbitals. Its numerical value, of course, is quite different, and we may expect a string scale near to the Planck length scale $\lp$: in particular, $\ls \sim 10 \lp \sim 10^{-32}$ cm for string models possibly including a realistic unification of all fundamental interactions \cite{16} (the string length $\ls$ could be larger, however, in models with large extra dimensions \cite{17,18}).

In any case, because of their finite size, quantum strings cannot occupy a vanishing volume, implying an upper limit to their energy density. In a cosmic context, on the other hand, the existence of a minimal volume suggests the existence of a minimal effective Hubble radius, $c/H \gaq \ls$, and thus a maximal value of the Hubble parameter and of the space-time curvature, $H \laq c/\ls$. If the curvature cannot blow to infinity because of such constraint, when a given space-time region reaches the limiting string value $ \sim \ls^{-1}$, its geometrical state can only evolve in two ways: it can either stabilize at that value (see e.g. \cite{19}), or start decreasing towards lower curvature states, after a bounce induced by appropriate  strong coupling effects ( see e.g. \cite{20}).

\begin{figure}[t]
\centering\includegraphics[width=5in]{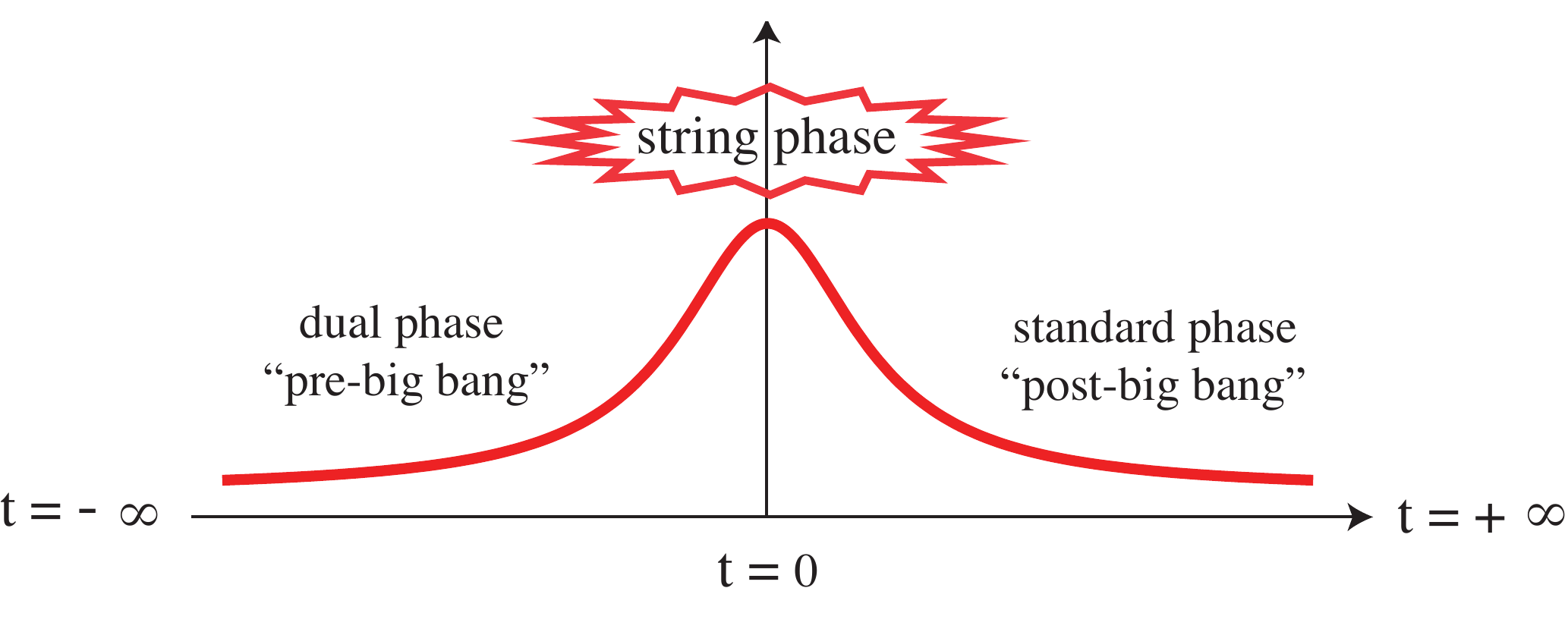}
\caption{Qualitative time evolution of the curvature scale and of the energy density in a typical example of pre-big bang string cosmology scenario. The Universe evolves starting from a flat, cold and empty state called ``string perturbative vacuum", asymptotically localized at $t=-\infty$. }
\label{f1}
\end{figure}

By combining  the existence of the dual symmetry and of a minimal length scale we can thus expect that a string-based model of the early Universe may complete the standard scenario by removing the curvature singularity, and extending the physical description of the cosmic space-time back in time, to infinity (as qualitatively illustrated in Fig. \ref{f1}). The big bang era may remain, but deprived of the standard role of initial singularity: it should correspond, instead, to a (possibly extended) phase of very high (but finite) maximal curvature -- the so called ``string phase" -- marking the transition between the growing curvature and decreasing curvature regime. 

Such a scenario, also inspired by pioneer papers on the duality properties of a cosmic ``string gas" distribution \cite{21,22}, was first explicitly derived from the cosmological equations of the string effective action and presented in detail in \cite{23}, and later developed in many papers (see e.g. \cite{24,25,26} for a review of further developments). In that context, as illustrated in Fig. \ref{f1}, the initial cosmological state is no longer localized at $t=0$ but is moved to the limit $t \ra -\infty$, and is assumed to coincide, asymptotically, with the so-called string perturbative vacuum (see in particular \cite{27,28} for a discussion of its properties): namely a flat, cold, empty initial state, drastically different form the hot, extremely curved and concentrated, explosive initial state proposed by the standard scenario.

It should be stressed, however, that the apparent specular symmetry between the pre-and post-big bang regime illustrated in the figure is expected to be broken (as we shall discuss below) by the asymmetric time evolution of the effective string coupling parameter $g_s$, controlled by the scalar dilaton field (not shown in the picture): the coupling, indeed, tends to zero as $t \ra -\infty$, while it grows in the opposite limit, and tends to become very strong, if not appropriately stabilized \cite{24,25}. Such an asymmetric behavior, by the way, is a crucial ingredient in the production of a realistic spectrum of scalar (curvature) perturbations via the axion-curvaton mechanism \cite{29,30,31,32,26}, and a possibly detectable cosmic background of relic gravitons, as recently discussed in \cite{33}.

\subsection{Scale-factor duality}


To give now a more technical illustration of the above results let us start with the gravi-dilaton string effective action, which -- to lowest order in the higher-derivative and string loop expansion -- can be written in the String frame and in $d+1$ space-time dimensions, as follows \cite{9,10}:
\beq
S=-{1\over 2\ls^{d-1}} \int_{\Om} \d^{d+1} x \,\sqrt{-g} \, \e^{-\phi} \left(R+ \pa_\mu\phi\,\pa^\mu \phi +V\right) + \int d^{d+1} x\, \rg \,{\cal L}_m .
\label{21}
\eeq
Here $\phi$ is the dilaton (the scalar string partner of the gravitational tensor field), and the last contribution to the action represents the possible contribution of gravitational sources: other fields arising from the low-energy  spectrum of string  states (like the Kalb-Ramond axion), radiation, a string gas distribution, etc. We have also included a possible dilaton potential $V$. The variation of the action with respect to the metric and to the dilaton gives, respectively, the field equations:
\bea
&&
G_{\mu\nu}+\nabla_\mu \nabla_\nu \phi  + {1\over 2} g_{\mu\nu} \left(\nabla \phi\right)^2
-g_{\mu\nu} \nabla^2 \phi- {1\over 2} g_{\mu\nu}V
=\ls^{d-1}\e^\phi ~T_{\mu\nu},
\label{22} \\ &&
2 \nabla^2 \phi-(\nabla \phi)^2+R+V- {\pa V\over \pa \phi}= \ls^{d-1} \e^{\phi} \sg, 
\label{23}
\eea
where $G_{\mu\nu}$ is the Einstein tensor, $\nabla_\mu$ the covariant derivative, $(\nabla \phi)^2 \equiv \nabla_\mu \phi \nabla^\mu \phi$ and $\nabla^2 \phi= \nabla_\mu \nabla^\mu \phi$. Finally, $T_{\mu\nu}$ the standard stress tensor defined as usual by the variation of the matter part of the action with respect to the metric, and $\sg$ is the scalar charge density representing the variational contribution with respect to $\phi$ of the matter sources, namely:  $\da_\phi (\rg \L) = -(1/2) \rg \,\sg \da \phi$.  

Let us consider, for the cosmological applications of this paper, a spatially flat and homogeneous (but not necessarily isotropic) gravi-dilaton background, whose     sources that can be represented in the perfect fluid form. We can thus set, in the synchronous gauge, 
\bea
&&
g_{\mu\nu}= {\rm diag} (1,- a_i^2 \da_{ij}), ~~~~~~~ a_i=a_i(t),
~~~~~~~ \phi=\phi(t),
\nonumber\\ &&
T_\mu\,^\nu= {\rm diag} (\r, -p_i \da_i^j), ~~~~~~~~ \r=\r(t),
~~~~~~~~ p_i=p_i(t), ~~~~~~~ \sg = \sg (t).
\label{24}
\eea
From the time and space components of the gravitational equations (\ref{22}) we then obtain
 \bea
 &&
 \fpu^2 -2  \fpu \sum_i H_i + \left( \sum_i H_i\right)^2- \sum_iH_i^2-V
 =2 \ls^{d-1}\,\e^\phi \r ,
 \label{25}
\\ &&
 \dot H_i -H_i \left(\fpu - \sum_k H_k\right) +{1\over 2} {\pa V\over \pa \phi} =   \ls^{d-1}\, \e^\phi \left( p_i- {\sg \over 2}\right),
 \label{26}
 \eea
where $H_1= \dot a_i/a_i$,  a dot denotes differentiation with respect to the comic time $t$, and the sum over $i,j,k$ runs from 1 to $d$. Form the dilaton equation (\ref{23}) we obtain:
 \beq
2 \fpp - \fpu^2 +2 \fpu \sum_iH_i- \sum_i\left(2\dot H_i+H_i^2\right)
- \left( \sum_i H_i\right)^2 +V- {\pa V\over \pa \phi}=\ls^{d-1}\, \e^\phi \sg. 
 \label{27}
 \eeq
 It should be noted that, in $d$ spatial dimensions, the combination $\ls^{d-1} e^ \phi $ plays the role of the effective gravitational coupling parameter, $16 \pi G \equiv 2 \lp^{d-1}$. The string coupling $g_s^2$ controlling the relative string-to-Planck length ratio is thus given, to this approximation, as
 \beq
 g_s^2 = \left(\lp\over \ls \right)^{d-1}= \left(\Mp\over \Ms \right)^{d-1} = e^\phi.
 \label{28}
 \eeq
 
 The above equations are invariant under the time reversal transformation $t \ra -t$, like the Einstein cosmological equations. However, they are different from the Einstein equations, and they are also invariant under the inversion of the scale factors, $a_i \ra a_i^{-1}$, provided the dilaton and the matter sources are also appropriately transformed. This symmetry, called scale-factor duality \cite{13,14}, has no analogue in general relativity and, for the particular case of $V=$ const and $\sg=0$, is represented by the following transformations:
 \beq
a_i \ra \ti a_i = a_i^{-1}, ~~~~~~~ \phi \ra \ti \phi= \phi- 2 \sum_{i} \ln a_i,  ~~~~~~~ 
\r \ra \ti \r = \r \prod_i a_i^2, ~~~~~~~ p_i \ra \ti p_i = - p_i \prod_k a_k^2.
\label{29}
\eeq
(Note that the sum and product symbols appearing here do not necessarily concern all the $d$ scale factors, but only those that have been inverted). By introducing the so-called ``shifted" variables  $\fb$, $\rb$, $\pb_i$, $\sgb$, defined by 
\beq
\fb=\phi- \ln \prod_i a_i= \phi- \sum_i \ln a_i, 
~~~~~
\rb= \r \prod_i a_i, ~~~~~ \pb_k= p_k \prod_i a_i,
 ~~~~~ \sgb= \sg \prod_i a_i,
 ~~~~~ i=1, \dots, d,
\label{210}
\eeq
the duality transformation (\ref{29}) can be conveniently rewritten also as follows:
\beq
a_i \ra a_i^{-1}, ~~~~~~~~~ \fb \ra \fb, ~~~~~~~~~ \rb \ra \rb, ~~~~~~~~~
\pb_i \ra - \pb_i, ~~~~~~~~~ \sgb=0.
\label{211}
\eeq
The above symmetry can be extended to the case of a non-trivial potential $V \not=$ const \cite{34,35,36} and of a non-vanishing scalar charge $\sg \not=0$ \cite{37}, provided both $V$ and $\sg$ depend on the dilaton through the shifted variable $\fb$ (and in that case the string model is described in general by a covariant but non-local effective action \cite{25,37}). 

\begin{figure}[t]
\centering\includegraphics[width=4in]{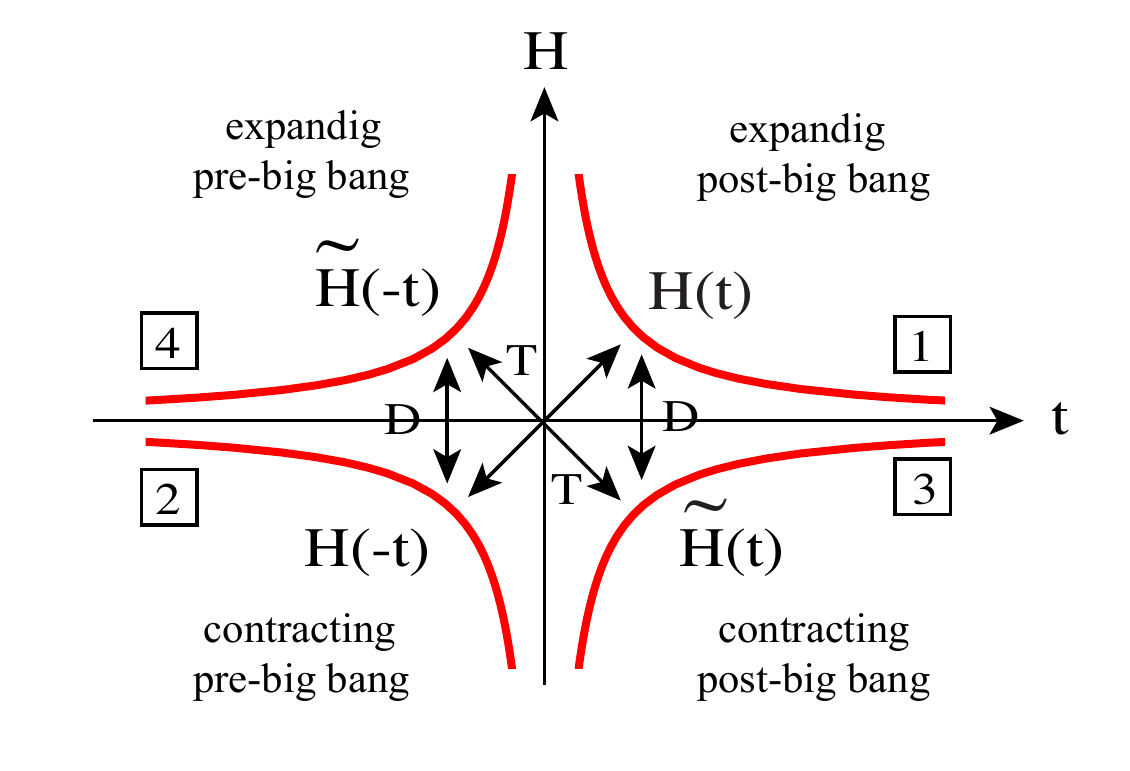}
\caption{Time evolution of the Hubble parameter in the four branches of the
isotropic vacuum solution (\ref{213}). The vertical arrows, labeled by ``D", represent duality transformations. The diagonal arrows, labeled by ``T", represent time reflections.}
\label{f2}
\end{figure}

According to the above symmetry, if the set of variables $S=\{a_i, \phi, \r, p_i\}$ represents an exact solution of Eqs. (\ref{25})-(\ref{27}) (with $\sg=0$, $V=$ const), then the transformed variables $S=\{\ti a_i, \ti \phi, \ti \r, \ti p_i\}$ of Eq. (\ref{29}) also represent a new (and physically different) exact solution of the same equations. To illustrate the existence of such duality-related branches of the string cosmology solutions let us first concentrate, for simplicity, on the case of $d$ isotropic dimensions, without sources and dilaton potential ($\r=p_i=\sg=V=0$). In that case, Eqs. (\ref{25})-(\ref{27}) are satisfied by the particular exact solution
\beq
a(t) = (t/t_0)^{1/\sqrt{d}}, ~~~~~~~~~~~~
\fb = - \ln (t/t_0), ~~~~~~~~~~~~~ t>0,
\label{212}
\eeq
where $t_0=$ const. Using duality and time-reversal transformations we can associate with this solution four different cosmological configurations,
\bea
&&
~~~~~~~~~~~~~~~~~\fbox{$1$} ~~~~~~~~~~~~~~~~~~~
{\rm time ~ reversal} ~~~~~~~~~~~~~~~~~~~~~~ \fbox{$2$}
\nonumber\\
&&
\{a \sim t^{1/\sqrt d}, ~~~ \fb \sim -\ln \, t\} ~~~~~ \Longleftrightarrow ~~~
\{a \sim (-t)^{1/\sqrt d}, ~~~ \fb \sim -\ln  (-t)\}
\nonumber\\
&&
~~~~~~~~~~~~~~~~~~~ \Updownarrow ~~{\rm duality}
~~~~~~~~~~~~~~~~~~~~~~~~~~~~~~~~~~~~~~{\rm duality} ~~ \Updownarrow
\nonumber\\
&&
\{a \sim t^{-1/\sqrt d}, ~~~ \fb \sim -\ln \, t\} ~~~ \Longleftrightarrow ~~~
\{a \sim (-t)^{-1/\sqrt d}, ~~~ \fb \sim -\ln  (-t)\}.
\nonumber\\
&&
~~~~~~~~~~~~~~~~~\fbox{$3$} ~~~~~~~~~~~~~~~~~~~
{\rm time ~ reversal} ~~~~~~~~~~~~~~~~~~~~~~ \fbox{$4$}
\nonumber\\
&&
\label{213}
\eea
The solutions [1] and [3] are defined for $t>0$, and disconnected by a curvature singularity from the solutions [2] and [4], defined for $t<0$. They have different (and complementary) kinematic properties. Two  branches  describe expansion ($H>0$), two branches contraction. 
Two branches (those defined for $t<0$) are characterized by growing curvature ($H^2$, or $|H|$, is growing in time), the other two branches ($t>0$) by decreasing curvature, as clearly illustrated in Fig. \ref{f2}. Finally, two branches ($t<0$) are accelerated (i.e. sign $\dot a=$ sign $\ddot a$), two branches ($t>0$) are decelerated (sign $\dot a=-$ sign $\ddot a$).

It is worth noticing, in particular, that because of the duality symmetry we can always associate to any given solution $a(t)$ of post-big bang type, evolving {\em from} the singularity, a new solution $a^{-1}(-t)$ of pre-big bang type, evolving {\em towards} the singularity. But it is also important to consider the dilaton behavior. In fact, the growing curvature solutions [2] and [4] (of pre-big bang type) are characterized by $\fbp>0$, while the decreasing curvature solutions [1] and [3] (of post-big bang type) are characterized by $\fbp<0$. A transition from the pre- to the post big bang regime must thus correspond to a transition from $\fbp>0$ to $\fbp<0$. 

The behavior of the pure dilaton field $\phi$ (and thus of the string coupling, see Eq. (\ref{28})), may be different, however. It may be appropriate, to this purpose, to rewrite the four branches (\ref{213}) in terms of the $\phi$ variable as follows:
\beq
a_\pm (\pm t)= (\pm t)^{\pm 1/\sqrt {d}}, 
~~~~~~~~~~~~~~~
\phi_{\pm} (\pm t)= 
(\pm \sqrt{d} -1) \ln (\pm t).
\label{214}
\eeq
It can be easily checked that among the branches at growing curvature, the solution $a_-(-t)$ -- corresponding to the curve [4] and describing an {\em expanding} pre-big bang configuration -- is associated to a { growing} dilaton and growing string coupling, while  
the solution $a_+(-t)$ -- corresponding to the curve [2] and describing a {\em contracting} pre-big bang configuration -- is associated to a {decreasing} dilaton and string coupling.  The choice of the initial state and of the initial value of the string coupling has thus implications also for the dynamical evolution of the  background geometry. Starting from the string perturbative vacuum (flat space-time and vanishing coupling, $H \ra 0$, $g_s \ra 0$), in particular, selects an expanding geometry (at least, to lowest order and in the string frame) for the pre-big bang regime. In that case, the transition to the post-big bang regime is expected to occur not only at high curvature, but also at strong coupling.

The above results can be easily generalized by including the presence of (homogeneous and isotropic) perfect fluid sources with barotropic equation of state, $p= \ga \r$, where $\ga=$ const (let us still assume $\sg=0=V$). For $\ga\not=0$ the cosmological equations (\ref{25})-(\ref{27}) are satisfied by the particular exact solution (in $d$ spatial dimensions):
\beq
a=\left(t\over t_0\right)^{2 \ga \over 1 +d \ga^2}, ~~~~~~~
\r=\rb a^{-d}= \r_0a^{- d(1+\ga)}, ~~~~~~~
\phi= {2(d\ga-1)\over 1+ d \ga^2} \ln \left(t \over t_0\right) +{\rm const},
\label{215}
\eeq
where $t_0$, $\r_0$ are positive integration constants. By applying the duality transformations (\ref{29}) and the usual time-reversal transformation we then obtain that the four different branches associated to this solution (disconnected by a curvature singularity at $t=0$) can be written as
\bea
&&
a_\pm(\pm t) \sim (\pm t )^{\pm {2 |\ga| \over (1 +d \ga^2)}}, 
~~~~~~~~~~~~
\r_\pm(\pm t) \sim (\pm t )^{\mp 2 {d|\ga|(1\pm |\ga|) \over (1 +d \ga^2)}}, 
\nonumber\\ && 
p_\pm= \pm |\ga| \r_\pm,
~~~~~~~~~~~~~ ~~~~~~~~~~~~~~~~~
\phi_\pm( \pm t) \sim -2{1 \mp d |\ga|\over 1 +d \ga^2} \ln (\pm t) + {\rm const}.
\label{216}
\eea
The positive-time branches are decelerated, with $a_+$ which describes expansion and $a_-$ which describes contraction (respectively, like the curves [1] and [3] of Fig. \ref{f2}). 
The negative-time branches are accelerated, of the pre-big bang type, with $a_+(-t)$ which describes contraction and $a_-(-t)$ which describes expansion (like the curves [2] and [4] of Fig. \ref{f2}). Again, as in the vacuum case, the solution $a_+(t)$, typical 
of the standard cosmological evolution, has a dual partner $a_-(-t)$
describing inflationary expansion. And, again, this dual ``complement" is associated to a dilaton field which (at least in the isotropic case) gives a string coupling which is always growing as $t \ra 0_-$, for any given equation of state:
\beq
g_s^2= \e^{\phi_-(-t)}= (-t)^{-2{1 + d |\ga|\over 1 +d \ga^2}} \ra + \infty,
~~~~~~~~~~ t  \ra 0_-.
\label{217}
\eeq

It may be instructive to consider the realistic example of a Universe with $d=3$  isotropic spatial dimensions, dominated by a radiation fluid with $\ga=1/3$. In this case, the expanding decelerated branch $\{ a_+(t), \phi_+(t), \r_+(t)\}$ of Eq. (\ref{216}) exactly reproduces the well known solution of the standard cosmological scenario,
\beq
a=t^{1/2}, ~~~~~~~~ \phi= {\rm const}, ~~~~~~~~ \r= \r_0\,a^{-4}, ~~~~~~~~
p=\r/3, ~~~~~~~~  t>0, 
\label{218}
\eeq
describing decelerated expansion, decreasing curvature, and frozen dilaton for $0 <t<\infty$:
\beq
\dot a >0, ~~~~~~~~~~~ \ddot a <0, ~~~~~~~~~~~  \dot H <0, 
~~~~~~~~~~~  \fpu =0.
\label{219}
\eeq
Through a dual inversion and a time reflection we obtain the associated inflationary partner $\{ a_-(-t), \phi_-(-t), \r_-(-t)\}$, i.e.:
\beq
a=(-t)^{-1/2}, ~~~~~~~~ \phi= -3 \ln (-t), ~~~~~~~~ \r= \r_0\,a^{-2}, ~~~~~~~~
p=-\r/3, ~~~~~~~~ t<0. 
\label{220}
\eeq
This solution is defined for $-\infty <t<0$, and describes a phase of accelerated expansion (driven by  the dilaton and by negative-pressure sources), growing curvature, and {\em growing dilaton}:
\beq
\dot a >0, ~~~~~~~~~~~ \ddot a >0, ~~~~~~~~~~~  \dot H >0, 
~~~~~~~~~~~  \fpu >0.
\label{221}
\eeq
It gives a typical example of pre-big bang evolution, with an almost trivial initial regime of nearly flat geometry and very small string coupling, asymptotically approaching (as already stressed) the string perturbative vacuum for $t \ra -\infty$. 


\subsection{$O(d,d)$ symmetry}


The scale-factor duality considered in the previous subsection is only a particular case of a more general symmetry of string theory, associated to  the invariance of the cosmological equations under global transformations of the pseudo-orthogonal  $O(d,d)$ group, and valid in general for time-dependent backgrounds with $d$ Abelian isometries \cite{34,35} (but it can be extended also to the non-Abelian case \cite{38,39}).

This global symmetry induces a non-trivial mixing of the metric and of the dilaton with other fields of the string spectrum, and can be illustrated, in the simplest case and to the lowest perturbative order, starting with the following string effective action:
\beq
S=-{1\over 2\ls^{d-1}} \int_{\Om} \d^{d+1} x \sqrt{-g} \, \e^{-\phi} \left(R+ \pa_\mu\phi\pa^\mu \phi +2\ls^{d-1}V-{1\over 12} H_{\mu\nu\a} H^{\mu\nu\a}\right) + \int d^{d+1} x\rg {\cal L}_m .
\label{222}
\eeq
With respect to the action (\ref{21}) we have added the contribution of the antisymmetric tensor field $H_{\mu\nu \a} = \pa_\mu B_{\nu\a}+ \pa_\nu B_{\a\mu}+ \pa_\a B_{\mu\nu}$, which is the ``field strength" of the NS-NS two-form $B_{\mu\nu}= - B_{\nu\mu}$ (also called Kalb-Ramond axion), appearing in the massless multiplet of bosonic string states \cite{9,10,11}. The field sources described by the matter part of the action are coupled in general to the metric, to the dilaton and to the two-form $B_{\mu\nu}$. Note that we have rescaled the dilaton potential in such a way that $V$ has now canonical dimensions of energy density. It should be stressed, finally, that the global $O(d,d)$ symmetry that we are considering can be implemented even including into the above action the higher derivative corrections (predicted by string theory in the high-curvature limit), to all orders in $\ls^2 = 2 \pi \ap$ \cite{40,41}.

Here we shall confine our discussion to the lowest perturbative perturbative order, and we will concentrate our attention on the class of homogeneous cosmological backgrounds which are isometric with respect to translations along $d$ spatial directions, so that we can choose a synchronous system  of coordinates where 
\beq
g_{00}=1, ~~~~~~~~~~~~~~~~~ g_{0i}=0= B_{0\mu},
\label{223}
\eeq
and the background fields $\{\phi, g_{ij}, B_{ij}\}$ are  functions of the cosmic time only. We shall also adopt a convenient matrix notation, defining the $d \times d$ matrix $G$ to represent the spatial part of the covariant metric tensor $g_{ij}$ (obviously, $G^{-1}$ will represent the controvariant components $g^{ij}$). In the same way, the spatial components of the two-form $B_{ij}$ are represented by the 
$d \times d$ matrix $B$. Using this formalism we can write, for instance, 
\beq
g^{ij} \dot g_{ji}= 
 {\rm Tr }\left(G^{-1}\dot G\right), ~~~~~~~
 H_{\mu\nu\a}H^{\mu\nu\a}= 3 H_{0ij}H^{0ij}=3 \dot B_{ij} (G^{-1} \dot B G^{-1})^{ij}=- 3{\rm Tr }\left(G^{-1}\dot B\right)^2,
 \label{224}
\eeq
and so on. Finally, we shall introduce the rescaled variable $\fb$, defined by
\beq
\e^{-\fb} =\ls^{-d}\int \d^dx \sqrt{|\det g_{ij}|}\, \e^{-\phi},
\label{225}
\eeq
useful to simplify the effective action (\ref{222}). Assuming that our background has spatial sections of finite volume, $(\int \d^d x \sqrt{|g|})_{t={\rm const}}=(\sqrt{|g|}V_d)_{t={\rm const}} < \infty$, and absorbing into $\phi$ the constant $\ln (V_d/\ls^d)$, we obtain
\beq
\fb= \phi -{1\over 2} \ln |\det G|, ~~~~~~~~~~
\fbp = \fpu -{1\over2} \Tr (G^{-1} \dot G)
\label{226}
\eeq
(we have used the identity $ |\det G|= \exp [{\rm Tr} \ln G]$). 

Summing up all terms arising from $\phi$, $R$ and $H_{\mu\nu\a}$, and neglecting a total derivative contribution, the effective action (\ref{222}) can then be expressed as a quadratic expression in the first derivative of the background field $\phi$, $G$, $B$, as follows:
\beq
S=-{\la_{\rm s}\over 2} \int \d t \,\e^{-\fb} \left[ \fbp^2
-{1\over 4} {\rm Tr }\left(G^{-1}\dot G\right)^2
+{1\over 4} {\rm Tr }\left(G^{-1}\dot B\right)^2 +V\right] + S_m.
\label{227}
\eeq
We are now in the position of discussing the invariance of the above action under global transformations of the  $O(d,d)$ group. 

To this purpose, let us first define a  $2d \times 2d$ matrix $M$, constructed from the spatial components of the metric and of the Kalb-Ramond field $B_{\mu\nu}$ as follows:
\beq
M=\begin{pmatrix}
G^{-1} & -G^{-1}B \\
BG^{-1} & G-BG^{-1}B \\
\end{pmatrix}. 
\label{228}
\eeq
Let us also introduce the invariant metric $\eta$ of the $O(d,d)$ group in the off-diagonal representation, 
\beq
\eta=\begin{pmatrix} 0 & I \\ I & 0 \\ \end{pmatrix}. 
\label{229}
\eeq
($I$ is the unit $d$-dimensional matrix). The matrix $M$ is symmetric, and belongs itself to the $O(d,d)$ group, since  $M^T \eta M =M \eta M=\eta$. It follows that 
\beq
{\rm Tr } \left(\dot M \eta\right)^2=
 {\rm Tr } \left( \dot M \dot M^{-1}\right)=
2  {\rm Tr }\left[\dot G^{-1}\dot G+  \left(G^{-1}\dot
B\right)^2\right]=
-2  {\rm Tr }\left(G^{-1}\dot G\right)^2 +2  {\rm Tr }\left(G^{-1}\dot
B\right)^2,
\label{230}
\eeq
where we have used the cyclic property of the trace. Using the above results, we find that the action (\ref{227}) can be recast in the form
\beq
S=-{\la_{\rm s}\over 2} \int \d t \,\e^{-\fb} \left[ \fbp^2
+{1\over 8} {\rm Tr } \left( \dot M \dot M^{-1}\right)+V\right] + S_m. 
\label{231}
\eeq

In this form, it can be easily checked that the kinetic part of the action is invariant under global $O(d,d)$ transformations preserving the shifted dilaton $\fb$, i.e. under the transformations
\beq
\fb \ra \fb, ~~~~~~~~~~~~~~~~~ M \ra \ti M= \Om^{T} M \Om, 
\label{232}
\eeq
where $\Om$ is a constant matrix satisfying $\Om^T \eta \Om =\eta$. Such an invariance is still valid for $V \not=0$ if the dilaton potential  is a constant, or is a function of the variable $\fb$ (or, more generally, it is a function of some $O(d,d)$ scalar formed with $M$ and $\fb$). 
Also, the scalar-factor duality  of the previous section can be recovered as a particular case of $O(d,d)$ transformation. In fact, let us consider a pure gravi-dilaton background with $B=0$, and a  global transformation generated by the particular matrix $\eta \in O(d,d)$. We have 
\beq
M = \begin{pmatrix}G^{-1} & 0\\ 0 & G \\ \end{pmatrix}, ~~~~~~~~~~~~~~~~
\ti M = \Om^{T} M \Om= \eta M \eta =
\begin{pmatrix}G & 0 \\ 0 & G^{-1} \\ \end{pmatrix},
\label{233}
\eeq
so that $G \ra \ti G = G^{-1}$, i.e. the considered transformation simply produces an  inversion of the spatial part of the metric. For a diagonal isotropic metric, in particular,  $G= - a^2I$, and the above equation is equivalent to the scale-factor duality transformation $a \ra \ti a = a^{1}$. 

The invariance under global $O(d,d)$ transformations can be preserved even including into the action the source contributions, provided their energy-momentum/charge/current densities, describing their coupling to the graviton/dilaton/axion fields, are appropriately defined and transformed \cite{36}.
To discuss this possibility it is convenient to write the cosmological equations following from the action (\ref{222}), and for a background satisfying the conditions (\ref{223}), in terms of $M$, $\fb$ and of the following ``shifted" variables associated with the sources:
\beq
\rb= \sqrt{-g}\, \r , ~~~~~~~~~~ \sgb= \sqrt{-g}\, \sg, ~~~~~~~~~~ {\overline \theta}^{ij} = \sqrt{-g} \,T^{ij}, ~~~~~~~~~~ {\overline J}^{ij}=\sqrt{-g} \,J^{ij},
\label{234}
\eeq
here $\r= T_{00}$, $\sg$ is the dilaton charge, $T^{ij}= T^{ji}$ is the spatial part of the energy-momentum tensor, and $J^{ij}=- J^{ji}$ is the spatial part of the tensor current density obtained by varying the matter action with respect to the NS-NS two form $B_{\mu\nu}$, namely:
\beq
\da_B \sqrt{-g} \L ={1\over 2} \sqrt{-g} \,J^{\mu\nu} \da B_{\mu\nu}.
\label{235}
\eeq
It is also convenient to work (as before) with the matrix variables $G$ and $B$, so that $\sqrt{-g} =|\det G|^{1/2}$, and  ${\overline \theta}^{ij}$ and ${\overline J}^{ij}$ are correspondingly represented by the $d\times d$ matrices ${\overline \theta}$ and ${\overline J}$. 

Let us suppose, for simplicity, that our sources are not directly coupled to the dilaton ($\sg=0$), but let us include a possibly non-trivial (and non-local) dilaton potential which can be expressed in terms of the shifted variable (\ref{225}) only, $V= V(\fb)$ \cite{25,36,37}. The set of cosmological equations following from the general action (\ref{222}), for the $d$-isometric background (\ref{223}), can then be written as follows (in units $2 \ls^{d-1}=1$):
\bea &&
\dot {\fb}^2 -2\ddot {\fb} -{1\over 8}\,{\rm Tr}\, (\dot M\eta)^2+{\pa V\over \pa \fb}-V =0,
\label{236}
\\ &&
\dot {\fb}^2 +{1\over 8}\,{\rm Tr}\, (\dot M\eta)^2-V = \rb \e^{\fb},
\label{237}
\\&&
{\d\over \d t} (M\eta \dot M) - \fbp (M\eta \dot M) = \e^{\fb} {\overline T},
\label{238}
\eea
where 
\beq
\overline T= \begin{pmatrix}-{\overline J},& - \overline \theta G+{\overline J}B
\\
G\overline \theta -B{\overline J},&
G\overline J G+B\overline J B -G\overline \theta B -B \overline \theta G
\\ \end{pmatrix} ,
\label{239}
\eeq
is the $2d \times 2d$ matrix representing the variational contribution of the matter action. Eq. (\ref{236}) corresponds to the matrix version of the dilaton equation, 
Eq. (\ref{237}) corresponds to the (0,0) component of the gravitational equations for the metric, and Eq. (\ref{238}) represent a combination of the spatial part of the of the gravitational equations and of the equation for the axion field $B_{\mu\nu}$. The above set of equations can also be obtained by varying with respect to $\fb$, $G$ and $B$ the action in matrix form (\ref{231}), by imposing the associated hamiltonian constraint $H=0$, and by combining the results in a final $2d \times 2d$ matrix form.

The string cosmology equations (\ref{236})-(\ref{238}) are exactly invariant under the global transformation defined by
\beq
\fb \ra \fb, ~~~~~~~~~~~ \rb \ra \rb, ~~~~~~~~~~~ M \ra \Om^T M\Om, ~~~~~~~~~~~
\overline T \ra \Om^T \,\overline T \Om,
\label{240}
\eeq
where $\Om^T \eta \Om =\eta$. This means that the $O(d,d)$ symmetry is preserved in the presence of sources provided the matrix $\overline T $ transforms exactly like $M$. A simple, but interesting, example of sources satisfying this property is a gas of classical, non interacting strings, minimally coupled to the background fields $\phi$, $G$, $B$ \cite{36}. But also an ordinary matter distribution composed of point-like particles and described  as fluid in the hydrodynamical approximation, is compatible with the given type of transformations. In that case, however, it should be stressed that the perfect fluid property is not in general invariant with respect to a generic $O(d,d)$ transformation \cite{36}, even if the axion background is vanishing. 

A simple illustration of this effect can be obtained by considering a perfect fluid source, characterized by $\overline \theta^i\,_j=-\pb \da^i_j$, $p=\ga \r$, $J^{ij}=0$, in a axion-vanishing, $B_{ij}=0$, isotropic metric background, $g_{ij}= - a^2 \da_{ij}$. The initial configuration is then represented by
\beq
M=\begin{pmatrix} G^{-1} & 0 \\ 0 & G \\ \end{pmatrix}, ~~~~~~~
G= - a^2 I, ~~~~~~ ~~
\overline T= \begin{pmatrix} 0 & \pb I  \\ -\pb I & 0 \\ \end{pmatrix}
\label{241}
\eeq
Let us consider, for simplicity, the case of  $d=2$ spatial dimensions, and let us apply to the above configuration the one-parameter $(O(2,2)$ transformation generated by the following $4 \times 4$ matrix \cite{42}
\beq
\Om(\a)={1\over 2} \begin{pmatrix} 1+c & s & c-1 & -s \\
-s & 1-c & -s & 1+c \\
c-1 & s & 1+c & -s \\
s & 1+c & s & 1-c \\ \end{pmatrix}, ~~~~~~
c\equiv {\rm cosh} \a, ~~~~ s\equiv{\rm sinh} \a,
\label{242}
\eeq
where $\a$ is a real parameter ranging from $0$ to $\infty$. It can be easily checked that $ \Om^T \eta \Om=\eta$. The transformed sources $\Om^T ~\overline T \Om$ still have $\overline J=0$, but are associated to a non-diagonal stress tensor, 
\beq
\ti{\overline \theta}^i\,_j =- \pb \tau^i\,_j, ~~~~~~~~~~~~~
\tau^i\,_j= \begin{pmatrix} c & -s  \\
-s & -c \\ \end{pmatrix},
\label{243}
\eeq
in a non-diagonal metric background 
\beq
\ti G = -{1\over 2 c a^2} 
\begin{pmatrix} c(1+ a^4)+a^4-1 & -s(1+a^4)  \\
-s(1+a^4) & c(1+a^4)-a^4+1\\ \end{pmatrix}.
\label{244}
\eeq
We should recall now that the stress tensor of a comoving fluid in a cosmic background geometry, including possible viscosity terms, can be parametrized in general as \cite{1,2}
\beq
\theta^i\,_j=-(p - \xi Z) \da^i_j +2 S \sg^i\,_j.
\label{245}
\eeq
Here $\xi$ and $S$ are, respectively, the bulk and shear viscosity coefficients, $Z= \nabla_\mu u^\mu$ is the expansion parameter, $u^\mu$ the geodesic velocity field of the fluid element, and
\beq
\sg^i\,_j= \nabla^i u_j -{Z\over d} \da^i_j
\label{246}
\eeq
is the traceless shear tensor in $d$ spatial dimensions. For the transformed metric (\ref{244}), on the other hand, we can easily find that $Z=0$, and that $\sg^i\,_j=H \tau^i\,_j$, with $H= \dot a/a$. Comparing Eqs. (\ref{243}), (\ref{245}) we are thus led to conclude that the transformed source can be consistently described as a pressure-less fluid with no bulk viscosity, and with shear viscosity proportional to the original pressure, i.e.
\beq
\ti {\pb}=0, ~~~~~~~~~
\ti \xi =0, ~~~~~~~~~ \ti S= -{p\over 2 H}= -{\ga \r\over 2H}.
\label{247}
\eeq
Such an effect of ``viscosity generation via duality" may have important  applications in a cosmological context, where the physical properties of the cosmic fluid can affect the evolution of the metric perturbations. 

Another important application of the duality symmetry (\ref{240}) is related to the cosmological effects produced by the presence of a non-vanishing axion background, possibly generated by the transformation of the generalized stress tensor (\ref{239}). Let us notice, to this purpose, that that by differentiating Eq. (\ref{237}), combining the result with Eqs. (\ref{236}), (\ref{238}), and using the $O(d,d)$ properties of $M$, we can write the general conservation equation for the energy density of the gravitational sources in matrix form as follows \cite{36}
\beq
\dot{\rb}+{1\over 4}\,{\rm Tr}\, (\overline T \eta M\eta \dot
M\eta)=0.
\label{248}
\eeq
If we now consider a perfect fluid with diagonal stress tensor,  $\overline \theta G= - \pb I$, evolving in a diagonal metric background with $G= -a^2 I$, and $G^{-1} \dot G= 2 HI$, we find that the above equation, written in terms of the physical (non-shifted) components, reduces to:
\beq
\dot \r+dH(\r+p) +{1\over 2} J^{ik}\dot B_{ik}=0.
\label{249}
\eeq

It follows that the antisymmetric tensor density $J^{ik}$,  in this equation, plays exactly the same role as that of an intrinsic ``vorticity tensor" appearing, for instance,  in the context of cosmological models with spinning fluid sources \cite{43,44}. Such a tensor,  on the other hand, is known to be the source of repulsive gravitational contributions,  possibly smoothing out the initial singularity  \cite{45}. This suggests that the same effect might also occur in the context of string cosmology models  with $B\not=0$. We shall conclude this subsection by presenting a very simple model confirming this possibility. 

Let us start with  an exact solution of the vacuum string cosmology equations, without sources and dilaton potential ($\overline T=0=V$), describing a background at constant dilaton, vanishing axion, and globally flat space-time geometry. Let us parametrize this trivial background solution with the so-called Milne coordinates as follows:
\beq
\d s^2= \d t^2- \left(t\over t_0\right)^2 \d x^2 - \d y^2- \d z_i^2, ~~~~~~~~~~~
\phi={\rm const}, ~~~~~~~~~~~B_{\mu\nu}=0,
\label{250}
\eeq
where $z_i$ are cartesian coordinates spanning an euclidean $(d-2)$-dimensional manifold, and $t_0$ is a constant parameter. The associated ``dual partner" solution, obtained through the scale-factor duality transformation (\ref{29}), is given by
\beq
\d s^2= \d t^2- \left(t\over t_0\right)^{-2} \d x^2 - \d y^2- \d z_i^2, ~~~~~~~~~
\phi=-2 \ln \left|t\over t_0\right| +{\rm const},  ~~~~~~~~~B_{\mu\nu}=0.
\label{251}
\eeq
This is also an exact solution of the vacuum string cosmology equations but, unlike the previous one, is characterized by a non-trivial 
geometry and a non-trivial dilaton. The curvature is growing in the negative time branch of the solution (describing accelerated expansion along the $x$ direction with $a(-t) \sim (-t)^{-1}$, $t<0$), decreasing  in the positive time branch (describing decelerated contraction  with $a(t) \sim t^{-1}$, $t>0$), and the two branches are separated by a singularity at $t=0$. 

Such a singular background can be regularized by an appropriate global transformation of the $O(2,2)$ group, in the sense that it can be transformed into another exact solution of the same equations characterized by a non-trivial evolution along two spatial directions (say, the $x$ and $y$ axes), and in which the negative and positive time branches are smoothly connected near the origin, without singularities \cite{42}. The same $O(2,2)$ transformation, applied to the trivial solution (\ref{250}), also provide a non-trivial but regular, duality-related solution which is smooth everywhere, and which asymptotically reproduces the regimes of linear expansion at $t \ra +\infty$ and linear contraction at $t \ra - \infty$, typical of the Milne metric. 

Let us concentrate our attention to the $\{x,y\}$ plane, where our initial solutions (\ref{250}), (\ref{251}) can be represented in terms of the $2\times 2$ matrices $G$ and $B$ as
\beq
B=0, ~~~~~ G_\pm=-\begin{pmatrix}a^2_\pm & 0  \\ 0 & 1 \\ \end{pmatrix}, ~~~~~
M_\pm= \begin{pmatrix} G^{-1}_\pm & 0  \\ 0 & G_\pm \\ \end{pmatrix}, 
~~~~~
a_\pm=\left|t\over t_0\right|^{\pm1}, ~~~~~ \fb= -\ln \left|t\over t_0\right|.
\label{252}
\eeq
The $\pm$ signs correspond, respectively, to the Milne metric (\ref{250}) and to its dual counterpart (\ref{251}). Let us apply to this initial background configuration  the duality transformation (\ref{232}) generated by the $O(2,2)$ matrix $\Om$ defined in Eq. (\ref{242}).  We obtain the set of transformed background fields $\{\ti G, \ti B, \ti \phi\}$,  defined by 
\bea
&&
\!\!\!\!\!\!\!\!\!\!
\ti \phi_\pm =-\ln \left[c\pm 1+ (c \mp 1)  (t/t_0)^2\right], ~~~~~
\ti G_\pm(\a) = -\begin{pmatrix}
{c\mp 1+ (c \pm 1) (t/t_0)^2 \over c\pm 1+ (c \mp 1) (t/t_0)^2} &
{s[1+(t/t_0)^2]\over c\pm 1+ (c \mp 1) (t/t_0)^2}\\
{s[1+(t/t_0)^2]\over c\pm 1+ (c \mp 1)(t/t_0)^2} & 1\\ \end{pmatrix},
\nonumber\\ 
&&
\!\!\!\!\!\!\!\!\!\!
\ti B_\pm(\a)=\begin{pmatrix}
0 &
{-s[1+(t/t_0)^2]\over c\pm 1+ (c \mp 1)  (t/t_0)^2}\\
{s[1+ (t/t_0)^2]\over c\pm 1+ (c \mp 1)  (t/t_0)^2} & 0\\ \end{pmatrix}
\label{253}
\eea
(recall that $c=\cosh \a$, $s=\sinh \a$, $\a=$ const). 

\begin{figure}[t]
\centering\includegraphics[width=3.7in]{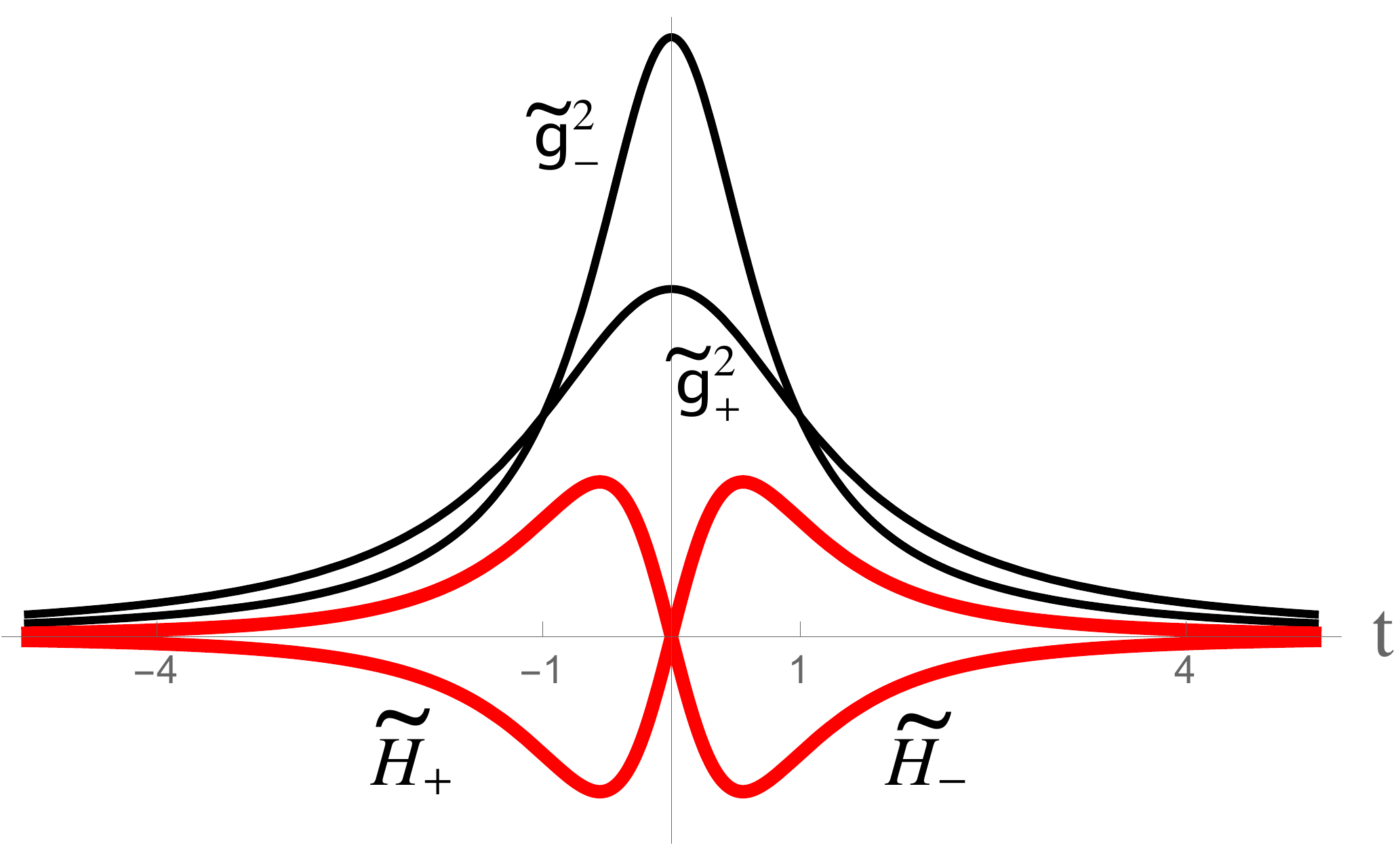}
\caption{Time evolution of $\ti H_\pm$ (bold red curves) according to Eq. (\ref{254}), and of $(\ti g_s^2)_\pm=\exp \ti \phi_\pm$ (thin black curves) according to Eq. (\ref{253}), for $\a=2$ and  $t_0=1$.}
\label{f3}
\end{figure}

It can be explicitly checked that these 
 new backgrounds fields exactly satisfy the string cosmology equations (\ref{236})-(\ref{238}) with $T_{\mu\nu}=0=V$. Unlike the initial backgrounds, however, the new ones  are non-trivial and all their curvature invariants are bounded, as well as the string couplings $(\ti g_s^2)_\pm= \exp(\ti \phi_\pm)$. They describe the  smooth evolution of a $(2+1)$-dimensional space-time from (anisotropic) contraction to expansion (or vice-versa), as can be shown by computing the ratio of change of the relative distance  between two comoving observers along the $x$ direction, which for the metric $\ti G_\pm$ is given by 
\beq
\ti H_\pm= V_{\mu\nu} \,n^\mu n^\nu =\pm { 4 c t\over t_0^2[c+1+(c-1)(t/t_0)^2][c-1 +(c+1)(t/t_0)^2]}.
\label{254}
\eeq
Here $V_{\mu\nu}= \nabla_{(\mu}u_{\nu)}$ is the so-called expansion tensor for a congruence of comoving geodesics $u^\mu$, and $n^\mu$ is a unit space-like vector along the $x$ direction, $n_\mu n^\mu=-1$, $ n_\mu u^\mu=0$. By taking $u^\mu= \da_0^\mu$ we have then, in the synchronous frame, $\ti H=-\dot g_{11}/2$. The plots of $\ti H_\pm$, together with those of $(\ti g_s^2)_\pm$, are given in Fig. \ref{f3} for the particular values $t_0=1$ and $\a=2$ (for $\a \gg1$ the two coupling parameters $(g_s^2)_+$ and $(g_s^2)_-$ tend to coincide).

In both cases, the boosted backgrounds $\ti M_+$ and $\ti M_-$ describe the evolution from a phase of growing curvature ad growing dilaton, to a phase of decreasing curvature and decreasing dilaton, as appropriate to a transition from a pre-big bang to a post-big bang configuration. Being  $(2+1)$-dimensional, and highly anisotropic,  this class of solutions does not seem appropriate to provide a realistic description of the present cosmological state. The situation may change, however, if the final configuration for $t \ra +\infty$ is modified by taking into account the back-reaction of the radiation produced in the transition from the accelerated to the decelerated regime. By including a third (initially trivial) spatial dimensions, one finds indeed that the radiation tends to become dominant, and tends to stabilize the dilaton and to isotropise the metric background, eventually turning the post-big bang contraction into a final phase of decelerated, isotropic expansion \cite{46}. Other examples of regular, homogeneous, isotropic (and possibly more realistic) backgrounds will be presented in the next Section. 


\section{Examples of smooth pre-big bang backgrounds}
\label{sec3}

\renewcommand{\theequation}{3.\arabic{equation}}
\setcounter{equation}{0}

In the previous section we have given an example of vacuum cosmological background ($T_{\mu\nu}=0$, $V=0$) regularized by the presence of a suitable axion field ($B_{\mu\nu}\not=0$). In this section we will illustrate the possibility of smooth solutions even in the absence of $B_{\mu\nu}$ contributions, and for spatially isotropic backgrounds, provided we include into the string cosmology equations (\ref{236})-(\ref{238}) an appropriate (non local) dilaton potential $V = V(\fb) \not=0$. Such a potential is expected to simulate the back-reaction of the string quantum-loop corrections, typically appearing in the limit of strong coupling and in the context of higher-dimensional space-time manifolds with extra dimensions of compact sections \cite{20,37}.

\begin{figure}[t]
\centering\includegraphics[width=3in]{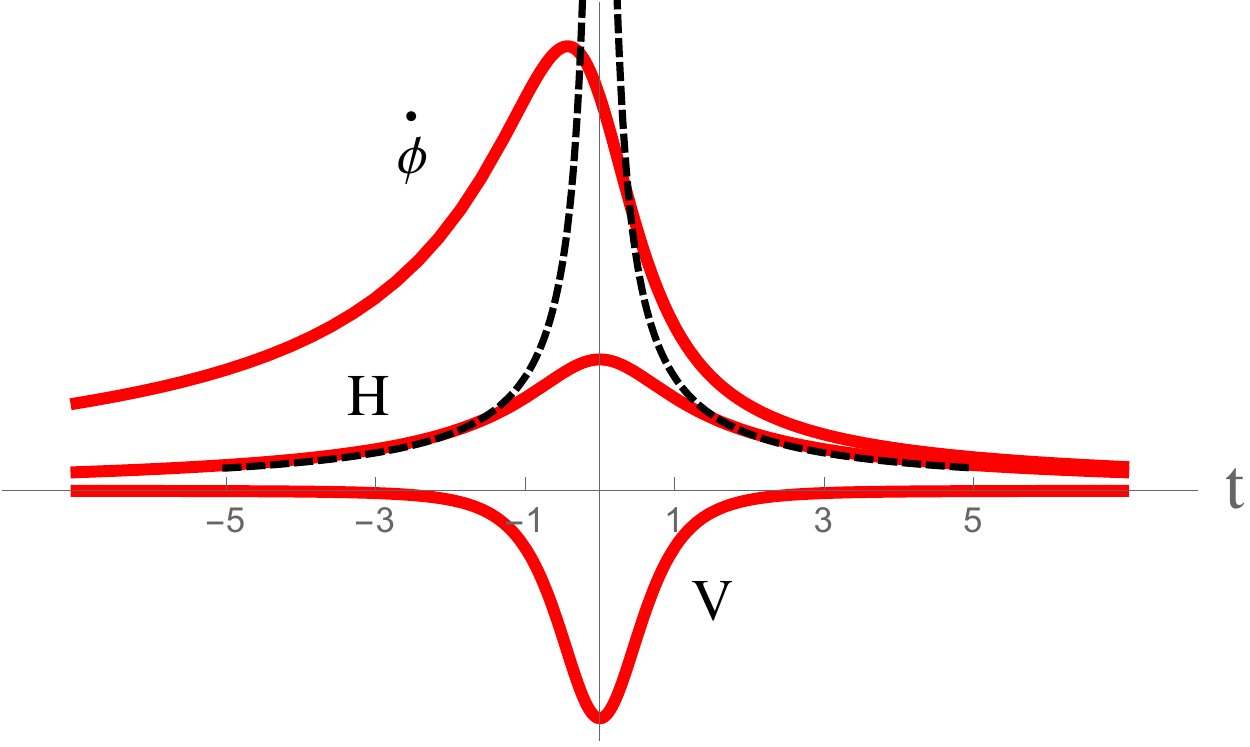}~~~~~~~~~~~~
\centering\includegraphics[width=2.3in]{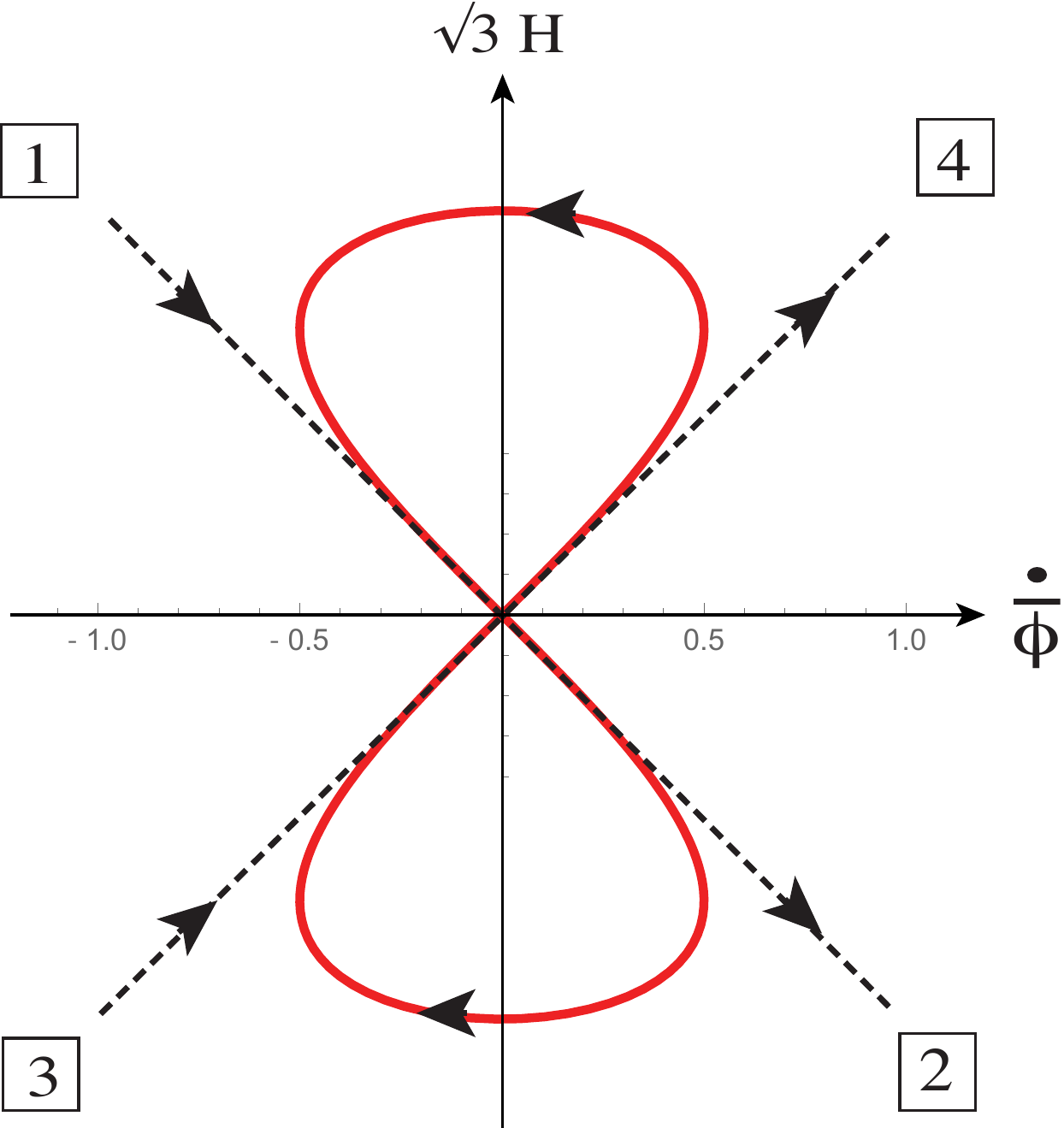}
\caption{The left panel shows the behavior of the curvature, of the dilaton kinetic energy and of the potential (\ref{31}) for the solution (\ref{32}). The dashed black  curves represent the (singular) expanding vacuum solutions of Eq. (\ref{213}), obtained for $V=0$. The right panel shows the trajectory of the solution (\ref{32}), and of its time-reversed partner, in the phase space spanned by $\dot \fb$ and $\sqrt{d} H$ (the dashed bisecting lines represent the four asymptotic vacuum solutions of Eq. (\ref{213})).  All curves are plotted for $t_0=1$, $V_0=1$, and $d=3$.}
\label{f4}
\end{figure}

Let us start with the simple vacuum case, $\r=p=\sg=0$, $B=0$, and with the following four-loop dilaton potential,
\beq
V(\fb)= -V_0 e^{4\fb}, ~~~~~~~~~~~~~~~~~ V_0>0,
\label{31}
\eeq
which is only a particular case of a more general class of ``$n$-loop potentials" (see e.g. \cite{20,47}). 
For that potential, and for the spatially flat, homogeneous and isotropic metric background with $G= -a^2 I$, the cosmological equations (\ref{236})-(\ref{238})  are exactly satisfied by the  following particular solution \cite{48}
\bea
&&
a(t)= a_0 \left[{t\over t_0} +\left(1+{t^2\over t_0^2}\right)^{1/2} \right]^{1/\sqrt d},
~~~~~~~~~~~
\fb= -{1\over 2} \ln \left[t_0 \sqrt{V_0} \left(1+{t^2\over t_0^2}\right)\right]+ {\rm const},
\nonumber \\ &&
\phi= \ln {\left[t/t_0+\left(1+{t^2/ t_0^2}\right)^{1/2}\right]^{\sqrt{d}}
\over \left(1+{t^2/ t_0^2}\right)^{1/2}} +{\rm const},
\label{32}
\eea
where $t_0$ and $a_0$ are positive integration constants. This is an exact   ``self-dual" solution, in the sense that   it satisfies $a(t)/a_0= a_0/a(-t)$, and is characterized by a bounded, ``bell-like" shape of the curvature scale and of the dilaton kinetic energy, as illustrated in Fig. \ref{f4} (left panel). Note that the potential dominates the background evolution in the high curvature, strong coupling limit $|t| \ra 0$, and becomes rapidly negligible as $t \ra \pm \infty$. 

The solution smoothly interpolates between the two expanding  pre-and post-big bang branches of the vacuum solution (\ref{213}, (\ref{214}): i.e., respectively, the branches called [4] and [1]  in Eq. (\ref{213}), which are reproduced by the new solution (\ref{32}) in the asymptotic limits $t \ra -\infty$ and $t \ra + \infty$, (and which are represented by the dashed curves in the left panel of Fig. \ref{f4}).  The right panel of Fig. \ref{f4} shows the trajectory in phase space of the solution (\ref{32}) and of its time reversal partner $a(-t)$, $\fb(-t)$. They smoothly evolve from an initial configuration of pre-big bang type ([4] or [2], with $\dot \fb >0$) to a final one of post-big bang type ([1] or [3], with $\dot \fb<0$). 

In the above solution, however, the dilaton $\phi$ keeps monotonically growing also in the post-big bang regime (indeed, $\phi \sim (\sqrt{d}-1) \ln t$ for $t \ra + \infty$). The string coupling $g_s \sim \exp \phi$ also monotonically grows, and this is not consistent, asymptotically, with the use of a tree-level effective action (higher order quantum corrections are needed, for consistency). In addition, when the dilaton growth is unbounded, the curvature might become singular in some frame (typically, in the so-called Einstein frame where the dilaton kinetic part of the action is canonically normalized and minimally coupled to gravity, see e.g. \cite{25}).

In more realistic examples such a dilaton growth is expected to be damped by the interaction with the matter/radiation post-big bang sources (and/or by the action of a suitable non-perturbative potential, possibly arising in the strong coupling regime). Let us notice, to this regard, that in the presence of sources the duality-invariant equations (\ref{236})-(\ref{238}) can be integrated exactly \cite{23} not only if $V=0$, as shown in Sect. \ref{sec2}, but, more generally, also if $2V= \pa V/\pa \fb$. Hence, also for a non-trivial two-loop potential given by
\beq
V(\fb)= -V_0 e^{2\fb}, ~~~~~~~~~~~~~~~~~ V_0= {\rm const}. 
\label{33}
\eeq
And it turns out that smooth solutions are in principle allowed if $V_0>0$ \cite{23}.

For an explicit illustration of this possibility it is convenient to work with a new (dimensionless) time-like coordinate $x$, and to represent the ``equation of state" of the sources, $\overline T =\overline T (\rb)$, in a differential form by introducing a suitable $2d \times 2d$ matrix $\Ga$ (possibly time-dependent). These two new variables $x$, $\Ga$ are defined by 
\beq
{dx\over dt} = {L\over 2} \rb, ~~~~~~~~~~~~~~~ \overline T = \rb {d\Ga \over dx},
\label{34}
\eeq
where $L$ is a constant length parameter (recall that we are using units in which $2 \ls^{d-1}=1$). We then find that the set of equations (\ref{236})-(\ref{238}), with the potential (\ref{33}), can be consistently integrated a first time to give two separate equations for $\fb$ and $M$:
\beq
\fb' =-{2\over D(x)} (x+x_0),
~~~~~~~~~~~~~~~~~
M\eta M'= {4\over D(x)} \Ga(x).
\label{35}
\eeq
Here $x_0$ is an integration constant, the prime denotes differentiation with respect to $x$, and 
\beq
D(x)= V_0L^2 +(x+x_0)^2 -{1\over 2}  {\rm Tr} \left(\Ga \eta\right)^2 .
\label{36}
\eeq

Let us now consider an isotropic and axion-less background with $B=0$ and $G=-a^2 I$, sourced by a perfect fluid with barotropic equation of state, $p/\r= \ga=$ const, $G \overline \theta - -\pb I$, and vanishing dilaton and axion charge density, $\sg=0$, $J=0$.  The simplest example of regular background can then be obtained for a ``dust" fluid source, $p=\ga=0$, which admit an (almost) trivial particular exact solution of Eqs. (\ref{34})-(\ref{36}) with $x_0=0$, $\Ga=0$, describing a globally flat space-time and a ``bell-like", time-symmetric evolution of the dilaton, sustained by a constant energy density \cite{49}:
\beq
a=a_0= {\rm const}, ~~~~~~~ \r=\r_0= {\rm const},  ~~~~~~~  p=0,  ~~~~~~~
\e^\phi= {\e^{\phi_0}\over 1 +\left(t/ t_0\right)^2}.
\label{37}
\eeq
The integration constants  $a_0$, $\r_0$, $\phi_0$, $t_0$ are related by the condition $\r_0 \exp \phi_0 = V_0 \exp (2 \phi_0)=4/t_0^2$. Such a solution acquires a less trivial representation in the Einstein frame \cite{20,47}, where the background smoothly evolves from $-\infty$ to $+\infty$ with a bell-like shape of the curvature, of the energy density and of the dilaton. In that frame, however, the pre-big bang regime corresponds to a phase of accelerated contraction, with a final bounce of the scale factor to the phase of (post-big bang) decelerated expansion.

The above example of smooth solution can be easily generalized to the case $\ga \not=0$, by considering for instance a radiation fluid with $\ga =1/d$ and/or its ``dual" partner with $\ga=-1/d$ (which can be phenomenologically interpreted as  a gas of ``winding" strings \cite{21,22}, or strings ``frozen" outside the horizon \cite{50,51}). The integration of the equations (\ref{35}) provides then regular solutions, interpolating between the pre- and the post-big bang regime, without singularities neither in the curvature nor in the string coupling \cite{23}. 

\begin{figure}[t]
\centering\includegraphics[width=3in]{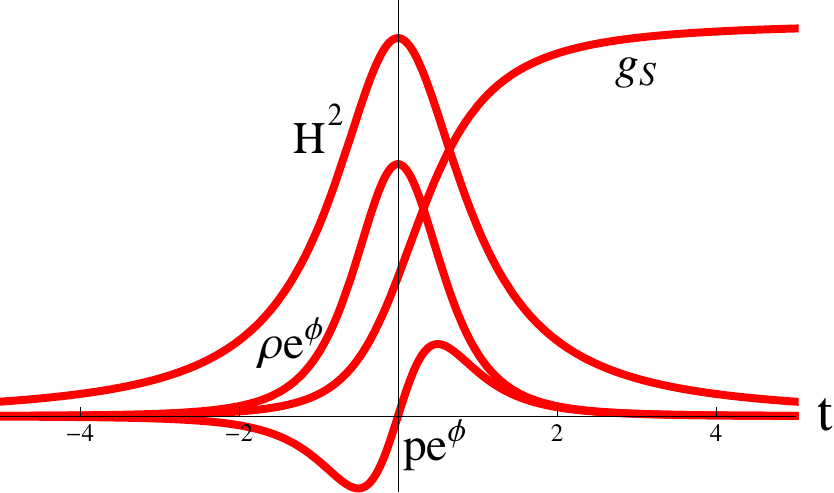}
\caption{Plot of the curvature scale, of the string coupling, of the effective energy density and of the effective pressure for the self-dual solution (\ref{39}). The curves are plotted for $d=3$, $L=1$, $x_1=1$, $\phi_0=0$, and $a_0=\exp(-2/3)$.}
\label{f5}
\end{figure}

It should be stressed, however, that the general equations  (\ref{34})-(\ref{36}) can be integrated exactly not only for fluids with barotropic equation of state ($\ga=$ const), but also for any given ratio $(p/\r)$ which is integrable function of  the time-like coordinate $x$.
An interesting example (motivated by the study of a string gas in rolling backgrounds \cite{52}) is the case in which $p/\r$ smoothly evolves from the value $\ga=-1/d$ at $t=-\infty$ to the value $\ga=1/d$ at $t=+\infty$ (thus connecting a final radiation phase to its initial dual partner), according to the law \cite{23}:
\beq
{p\over \r}={1\over d} {x\over \sqrt{x_1^2+x^2}} 
\label{38}
\eeq
($x_1$ is a free parameter). By integrating Eq. (\ref{34}) to obtain $\Ga$, and limiting (as before) our discussion to the case of isotropic backgrounds with $B=0$ and $\sg=0$, a second integration of the equations (\ref{35}) (with the simplifying choice of the free parameters $x_0=0$ and $x_1^2 =L^2 V_0$), 
leads to the following particular exact solution:
\bea
&&
a=a_0 \left(x+ \sqrt{x^2+x_1^2}\right)^{2\over (d-1)},
~~~~~~~~~~~~~~~~
\e^\phi=a_0^d \e^{\phi_0}\left(1+{x\over \sqrt{x^2+x_1^2}}\right)^{2d\over(d-1)}, 
\nonumber\\ &&
\r \e^\phi= {d-1\over dL^2} \e^{2\phi_0} \left(x^2+x_1^2\right)^{-{(d+1)\over(d-1)}}, 
~~~~~~~~
p\e^\phi= {d-1\over d^2L^2} \e^{2\phi_0}x \left(x^2+x_1^2\right)^{-{(3d+1)\over2( d-1)}}, 
\label{39}
\eea
where $a_0$ and $\phi_0$ are integration constants. The smooth and bouncing behavior of this solution is illustrated in Fig. \ref{f5}.

The above solution is self-dual, in the sense that $\fb(x)=\fb(-x)$, $\rb(x)=\rb(-x)$, and
\beq
\left[a(x)\over a_0 x_1^{2/(d-1)}\right]=
\left[a(-x)\over a_0 x_1^{2/(d-1)}\right]^{-1}
 \label{310}
 \eeq
(with an appropriate choice of the integration constant $a_0$ it is always possible to set to $1$ the fixed point of scale-factor inversion). We also note that, asimptotically, we have $a \sim (-x)^{-2/(d-1)} \sim \rb \sim {\d x/ \d t}$ for $x \ra - \infty$, and $a \sim x^{2/(d-1)}\sim  1/{\rb} \sim {\d t/ \d x}$ for $x \ra + \infty$. 
In the asymptotic limit we can then easily re-express the solution in terms of the cosmic time 
coordinate, to find that it describes a monotonic evolution from an initial state of accelerated expansion and growing dilaton, associated to  negative-pressure matter, 
\beq
x \ra - \infty ~~~~ \Longrightarrow ~~~~
a \sim (-t)^{-{2\over d+1}}, ~~~~ \e^\phi \sim (-t)^{-{4d\over d+1}}, ~~~~ p=-{\r \over d}, 
\label{311}
\eeq
towards a final, radiation-dominated state of decelerated expansion 
and asymptotically frozen dilaton,
\beq
x \ra  +\infty ~~~~~~~ \Longrightarrow ~~~~~~~ 
a \sim t^{{2\over d+1}}, ~~~~~~~ \e^\phi \sim{\rm const}, ~~~~~~~ p={\r\over d}. 
\label{312}
\eeq
It follows that our solution (\ref{39}) smoothly interpolates between the pre-big bang and post-big bang  configurations obtained  {\em without} the contribution of the dilaton potential, and described, respectively, by the solutions  (\ref{220}) and (\ref{218}) (in $d=3$ spatial dimensions) of the string equations.  Hence, as  in the previous case, the smoothing out of the tree-level singularity, and the appearance of bouncing transition, is a consequence of the effective potential (\ref{33}). 

Let us note, finally, that the smooth, interpolating solution (\ref{39}) is in principle associated to a dual partner which corresponds to the same time-dependent equation of state as that of Eq. (\ref{38}), but with the opposite sign. It describes a background in monotonic contraction, with the sources evolving from  positive to  negative pressure. The combination of the two types of solutions may suggest a  $(1+d+n)$-dimensional, anisotropic scenario, in which the $d$  ``external" spatial dimensions are isotropically and monotonically expanding, while the $n$ ``internal" spatial dimensions are isotropically and monotonically contracting. Such a scenario may simultaneously include both inflation and dynamical dimensional reduction (see e.g. \cite{25}). 


\section{Is our Universe self-dual?}
\label{sec4}


We have shown in the previous sections that regular, self-dual solutions of the effective string cosmology equations are in general allowed, and are possible candidates to describe the very early cosmological evolution, completing (in a symmetric way)  the picture of the standard scenario. 

However, in a realistic model of our Universe, such a self-duality symmetry may be expected to be (al least softly) broken, because of three main effects: $(i)$ the appearance of strong-coupling (i.e., higher-loop) contributions to the string effective action; $(ii)$ the consequent stabilization of the shrinking extra dimensions and of the dilaton; $(iii)$ the amplification of the quantum fluctuations of the vacuum and the consequent effects of particle production, associated with the bounce of the scale factor. 

I have no room in this work to present a detailed discussion of these effects, but let me stress that the typical imprint of a (broken or not) duality symmetry of the cosmological evolution is the amplification of  (scalar and tensor) metric perturbations with a strongly blue (i.e., growing with frequency) primordial spectrum (as pointed out long ago, see e.g. \cite{49,53,54,55,56,56a}). This  effect has two important consequences.

Concerning scalar metric fluctuations, a steep and growing primordial spectrum is in obvious conflict with present data, and has to be efficiently suppressed at all (large) scales relevant to the present CMB observations. This is possible, provided the parameters controlling the cosmic pre-big bang dynamics satisfy appropriate constraints \cite{33}. In addition, an appropriate (slightly red) spectrum of adiabatic curvature perturbations, responsible for the observed CMB anisotropy, has to be efficiently produced through the axion-curvaton mechanism \cite{29,30,31,32}. This also can be obtained for a restricted class of pre-big bang backgrounds \cite{24,25,26,33}, and the restrictions imposed to match the observed spectral behavior can break the perfect duality symmetry of the whole cosmological scenario. 

Even imposing such constraints (and thus obtaining scalar perturbations compatible with observational data) we are left however with tensor perturbations whose properties are somewhat ``anomalous" with respect to those of the standard inflationary scenario. In fact, their amplification during the pre-big bang phase leads to the production of a primordial stochastic background of relic gravitational waves peaked at high frequency (typically, in the GHz range), and with a large, possibly detectable intensity in the sensitivity band of present (or near-future) interferometric antennas (see e.g. \cite{33,57} for a recent discussion). The spectral properties of these cosmic gravitons are uniquely and directly correlated to the kinematic behavior of the gravi-dilaton background in the primordial pre-big bang regime. Their detection would thus provide direct (and unique) information on the possible dual symmetry of our Universe, and/or on its possible degree of breaking.

\section{Conclusion}

To conclude this short review, let me take the liberty of presenting a ``biological"  analogy -- which has no claim of scientific validity, of course, but which I find  suggestive -- between the big bang and the childbirth moment.

The big bang marks the beginning of our Universe in the same way as the childbirth marks the beginning of a new human life. The childbirth, however, is not a sudden and spontaneous event: it is prepared by a long period  of (nine months) pregnancy. In the same way, we may expect that the big bang epoch, also, should  be prepared, in some way. String cosmology suggests that it might be prepared by a long epoch of pre-big bang evolution at growing curvature, growing density, growing coupling, preceding the final ``delivering" explosion.

\section*{Acknowledgement}
It is a great pleasure to thank Gabriele Veneziano for many years of exciting and very fruitful collaboration on various aspects of cosmology and string theory, producing most of the results reported in this paper.
The author is supported in part by INFN under the program TAsP: {\it ``Theoretical Astroparticle Physics"}, and by the research grant number 2017W4HA7S {\it ``NAT-NET: Neutrino and Astroparticle Theory Network"}, under the program PRIN 2017 funded by the Italian Ministero dell'Universit\`a e della Ricerca (MUR).

\end{document}